\documentclass[showpacs,aps,pra,twocolumn,amsmath,amssymb]{revtex4-1}
\usepackage{graphicx}
\usepackage{dcolumn}
\usepackage{bm}
\usepackage{epsf}
\usepackage{amsmath}
\usepackage{color}

\newcommand{\la}{\langle}
\newcommand{\ra}{\rangle}
\newcommand{\be}{\begin{equation}}
\newcommand{\ee}{\end{equation}}
\newcommand{\bea}{\begin{eqnarray}}
\newcommand{\eea}{\end{eqnarray}}

\begin{document}

\title{Propagation properties and limitations on the attainable entanglement in a driven harmonic chain}

\author{Fernando  Galve}
\affiliation{IFISC (UIB-CSIC), Instituto de Fisica Interdisciplinar y Sistemas Complejos, UIB Campus, E-07122 Palma de Mallorca, Spain
}

\date{\today}

\begin{abstract}
The limitations on the production and profitability of entanglement in a harmonic chain under strong driving are considered. We report on the limits of attainable entanglement for a given set of squeezings of the eigenmodes, showing that the higher the entanglement the more oscillatory and thus less easy to profit from. We also comment on propagation properties of entanglement, discussing the role of fast rotating terms and illustrating several issues with the example of a sudden switch of the coupling. 
\end{abstract}

\pacs{03.67.Bg}

\maketitle

\section{Introduction}

Entanglement is recognized not only as a striking feature of quantum mechanics, but also as an important resource for quantum information tasks. Its creation and manipulation will be a must if we are to profit from the quantum speed up of some information protocols that are not available in the classical world. Many body systems, studied under the light of quantum information, have provided \cite{ami07} new tools for the condensed matter community, while they can be regarded as interesting devices for the production and manipulation of entanglement. 

A particular example of such systems is the harmonic chain, whose entanglement properties have been extensively studied \cite{ami07,aud02,anders07}. In addition to the static properties of entanglement in the chain, its production and manipulation starting from a ground or thermal state have been analyzed in \cite{eis04,njp}, where it was shown that parametric changes in the coupling lead to production of long distance entanglement. This production can be optimized \cite{gal2009} and is robust against realistic dissipation from hot environments \cite{gal2010}. Furher, the harmonic chain can be mapped into experiments with nanomechanical resonators \cite{rou01} and ion crystals in multitrap arrangements \cite{wak92}, and has the exceptional features of being exactly solvable and having an exact measure of entanglement \cite{vid02}.

In this paper we present a detailed analysis of the entanglement properties of a harmonic chain with nearest neighbor time dependent coupling and give expressions for the maximum attainable entanglement between distant oscillators when the chain is initially in the ground state. These expressions are obtained for weak and strong coupling regimes, and for any amount of squeezing in the eigenmodes. It is shown that a high amount of squeezing can lead to higher entanglement but at the expense of it being strongly oscillatory. This is explained in terms of the optimality of angular relations between different eigenmodes, and how they affect the usefulness of the squeezings to produce entanglement. We show how these optimal relations shrink to smaller sets for higher squeezing or coupling, and quantify this transition. The attainable entanglement is explicitly calculated for the simple case of a sudden switch of the coupling.

The propagation properties in the chain have been studied for example in \cite{njp}. We show here that reasoning in terms of the group velocity in the continuous limit yields correct results for the speed of entanglement, as well as insight on the dispersive properties of the chain as a medium for excitations. We relate the importance of fast rotating terms to the strength of the coupling, as well as to the ability to produce entanglement with an optimal modulation, and compare the harmonic chain to the anisotropic XY spin chain where the dispersion relation has more parameter freedom.

\section{Harmonic chain}
The system under consideration is a closed chain of $N$ harmonic oscillators with identical frequency $\omega_0$ and harmonic coupling between nearest--neighbor $c(t)$:
\begin{equation} \label{eq1}
{H}=\frac{1}{2}\sum_{n=1}^N \Big({p}_n^2+\omega_0^2{q}_n^2+c(t)({q}_{n+1}-{q}_n)^2 \Big) \ .
\end{equation}
(we will use $m=\hbar=1$ throughtout the paper).
To diagonalize the Hamiltonian  \eqref{eq1}, we introduce the normal mode coordinates $Q_l$ and $P_l$ via,
\begin{equation}
\label{eq2}
{q}_n=\frac{1}{\sqrt{N}}\sum_{l=1}^N e^{2\pi i ln/N} Q_l, \hspace{.6cm} {p}_n=\frac{1}{\sqrt{N}}\sum_{l=1}^Ne^{2\pi i ln/N}P_l \ .
\end{equation}
 The Hamiltonian of the chain 
\begin{equation} \label{eq3}
{H}=\frac{1}{2}\sum_{l=1}^N \left({P}_l{P}_l^\dagger+\omega_l^2{Q}_l {Q}_l^\dagger\right) \ ,
\end{equation} 
is then that of a set of independent oscillators with time--modulated frequencies, $\omega^2_l(t)=\omega_0^2+4 c(t)\sin^2(\pi l/N)$ (we have used the property, $Q_{-l}=Q_l^\dagger$, $P_{-l}=P_l^\dagger$).
 The linear Heisenberg  equations of motion for $P_l(t)$ and $Q_l(t)$  can readily be written down as:
\begin{eqnarray}
Q_l(t)&=&Q_l(0)Y_l(t)+P_l^\dagger(0)X_l(t)\\
P_l(t)&=&\frac{d}{dt}Q_l(t)=Q_l(0)\dot{Y}_l(t)+P_l^\dagger(0)\dot{X}_l(t)
\end{eqnarray}
where the functions $X_l(t)$ and $Y_l(t)$ are solutions to the classical equation of motion $\ddot{x}=-\omega_l(t)^2x$ with initial conditions $Y_l(0)=1$, $\dot{Y}_l(0)=0$, $X_l(0)=0$, $\dot{X}_l(0)=1$.
We will restrict ourselves here to Gaussian states, whose complete information is contained in the symmetric $2 N \times 2N$ covariance matrix $\Gamma$. Its elements are defined as: $\Gamma_{q_nq_m} = 2 \mbox{Re} \la q_nq_m \ra$, $\Gamma_{q_np_m} = 2 \mbox{Re} \la q_np_m \ra$ and $\Gamma_{p_np_m} = 2 \mbox{Re} \la p_np_m \ra$.  The first moments are of no relevance to the entanglement properties and we can drop their description from now on.
If we assume that the chain is prepared in the \textit{ground state} of the noninteracting Hamiltonian, the time-evolved elements of the covariance matrix can be written in a simple way:
\be
\label{eq4}
\la{q}_n{q}_m\ra=\frac{1}{N}\sum_{l=1}^Ne^{2\pi i l(n-m)/N}\la{Q}_l{Q}_l^\dagger\ra \ ,
\ee
where we have used $\omega_0\Gamma_{q_nq_m}(0) = \Gamma_{p_np_m}(0)/\omega_0 = \delta_{n,m}$, $\Gamma_{q_np_m}(0) =0$ and $\la{Q}_r{Q}_s^\dagger\ra=\la{Q}_s{Q}_s^\dagger\ra\delta_{r,s}$. The latter equality holds only in the case where the eigenoscillators were uncorrelated at the beginning, and because they never interact, will always be. Similar expressions are obtained for $\la{q}_n{p}_m\ra$ and $\la{p}_n{p}_m\ra$. The problem has been reduced to finding the second moments of a set of independent oscillators with modulated frequency. 

It is known that a time--dependent  oscillator is squeezed when its frequency is changed nonadiabatically, while an adiabatic transformation leads to  no squeezing \cite{jan87}. The frequency modulation effects a Bogoliubov transformation in the ladder operators of each oscillator: $a\rightarrow \mu a+\nu a^\dagger$ which keeps the canonical commutation relations $|\mu|^2-|\nu|^2=1$. They can be parameterized as $\mu=\cosh r$ and $\nu=-e^{i2\theta}\sinh r$, thus yielding
\begin{subequations}
\label{eq5}
\be
\la{Q}_l{Q}_l^\dagger\ra=\frac{1}{2\omega_l}(e^{-2r_l}\cos^2\theta_l+e^{2r_l}\sin^2\theta_l)\ ,
\ee
\be
\la{P}_l{P}_l^\dagger\ra=\frac{\omega_l}{2}(e^{2r_l}\text{cos}^2\theta_l+e^{-2r_l}\text{sin}^2\theta_l)\ ,
\ee
\be
\la{Q}_l{P}_l^\dagger\ra=\text{sinh}(2r_l)\text{sin}\theta_l\text{cos}\theta_l \ .
\ee
\end{subequations}
for the quadratures of the eigenmodes. The time dependence of the eigenfrequencies $\omega_l$, the squeezing parameters $r_l$ and the squeezing angles $\theta_l$
 are all controlled by the linear coupling coefficient $c(t)$.

\section{Attainable entanglement}
The logarithmic negativity \cite{vid02} gives an exact quantification of the bipartite entanglement between oscillators $n$ and $m$,
\begin{equation}
\label{eq6}
E_N=\max(0,-\log_2(|\nu_-|).
\end{equation}
 Here,  $\nu_-$ is the smallest symplectic eigenvalues of the reduced and partially transposed covariance matrix of the two oscillators. 

We will consider in this study only opposite oscillators, $m=N/2+n$, but any other pair could as well be easily described with our approach.
 Opposite pairs have the largest possible distance in the chain, and it can be shown that they exhibit the largest value of entanglement, due
 to interference between signals coming through the two branches of the chain. In addition, their exponent in the quadratures becomes $e^{2\pi il(n-m)/N}=(-1)^l$,
 thus distinguishing even and odd normal modes. Due to the translational invariance, the logarithmic dependence of pairs of oscillators doesn't
 depend on their position $n$ but only on their relative distance $n-m$.

The symplectic eigenvalues of the $4 \times 4$ reduced covariance matrix $\Gamma_{n,m}$ can be written explicitly using local symplectic invariants \cite{invariants},
 which remain unchanged by operations on only one of the oscillators. A reorganization of the invariants for opposite oscillators reveals that the symplectic eigenvalues
 can be expressed as sums of a unique quantity $A_{l,m}$:
\begin{eqnarray}
\nu_-&=&\frac{\sqrt{2}}{N}\sqrt{x-\sqrt{x^2-4y}}\label{eq_eigen}\\
x&=&\left(\sum_{l,\text{odd}}\sum_{m,\text{even}}+\sum_{l,\text{even}}\sum_{m,\text{odd}}\right) A_{l,m}\\
y&=&\left(\sum_{l,\text{odd}}\sum_{m,\text{odd}} A_{l,m}\right)\left(\sum_{l,\text{even}}\sum_{m,\text{even}} A_{l,m}\right)\\
A_{l,m}&=&\frac{\omega_m}{\omega_l}(e^{-2r_l}\cos^2\theta_l+e^{2r_l}\sin^2\theta_l)\times\nonumber\\
&&(e^{2r_m}\cos^2\theta_s+e^{-2r_m}\sin^2\theta_m)-\nonumber\\
&&\sinh 2r_l\sinh 2r_m\sin2\theta_l\sin2\theta_m
\end{eqnarray}
From eq.(\ref{eq_eigen}) it is clear that the entanglement between opposite oscillators will be highest when $4y<<x^2$.
 This implies minimization of $\text{A}_{\text{odd,odd}}$ {\it and} $\text{A}_{\text{even,even}}$ and maximization of 
$\text{A}_{\text{odd,even}}$ {\it or} $\text{A}_{\text{even,odd}}$ for all eigenmodes. Clearly, this is almost intractable and a numerical analysis is here required. 

Of course the dynamical evolution given by the time function $c(t)$ dictates the sets $\{r_l(t)\}$ and $\{\theta_l(t)\}$.
 However, when modulation is over, and a finite constant coupling $c$ is kept, the squeezings do not change, but every angle
 oscillates with its own frequency $\omega_l$. We study next which combinations of angles are optimal, in terms of production
 of entanglement betwen opposite oscillators, for a given set of generated squeezings.

\subsection{Optimal angles at moderate squeezings and small coupling}
The eigenfrequencies of the normal modes are symmetric with respect to $N/2$. That means that $\omega_{N-i}=\omega_i$, $\omega_{N/2}$
 is the greatest, and $\omega_N=\omega_0$ which we set to 1. With that in mind it is clear that we only need to analyze the behavior
 of half the number of eigenmodes. Since opposite oscillators have a separated odd-even structure in the covariance matrix the first
 nontrivial case for a closed chain is having four oscillators. In that case the number of eigenmodes, and of angles, to be analyzed is 2. 

The regime $c<<\omega_0^2$ and moderate squeezing was studied in \cite{gal2009}; in this regime $\omega_m/\omega_l\sim 1$. There it 
was shown that there exists optimal relations for the angles:
\begin{eqnarray}
\label{oldAngles1}
\theta_l-\theta_m&=&2k\pi/2 \text{\ \ \ \ \ \ \ \ \   ,\ \  $l+m$ even}\\
\label{oldAngles2}
\theta_l-\theta_m&=&(2k+1)\pi/2 \text{\ \ ,\ \  $l+m$ odd}
\end{eqnarray}
with $ k \in\mathbb{Z}$.

We have repeated such analysis for different number of oscillators, and obtained the same result.
 Hence maximum entanglement is achieved when both even and odd oscillators synchronously achieve 
those relations for the angles. The limit on attainable logarithmic negativity is thus:

\begin{equation}
\label{eq8}
E_N^{max}=\!-\frac{1}{2}\text{log}_2\!\left[\!\left(\frac{2}{N}\!\!\sum_{l,odd}e^{-2r_l}\right)\!\!\left(\frac{2}{N}\!\!\sum_{m,even}e^{-2r_m}\right)\!\right]\!.\!
\end{equation}
Entanglement for opposite oscillators is only nonzero for nonvanishing squeezings. 

The optimal angular relation is modified (and thus the validity of eq.\eqref{eq8} ) when the coupling or the squeezings increase. A self-consistency argument (see Appendix \ref{App:Validity}) shows that the optimal relations for the angles is valid up to:
\begin{equation}
\label{self-cons}
\frac{c}{4\omega_0^2}\sum_le^{2r_l}\sim \sum_le^{-2r_l}.
\end{equation}
In next section we will show that for higher squeezings and coupling the optimal angles shrink to a smaller set of values. This immediately means that for a set of strong squeezings $\{r_s\}$ it will much harder to obtain dynamically (by independent rotation of each eigenmode at frequency $\omega_s$) the optimal combination of angles, and thus entanglement will be higher but much more oscillatory. This seems to be a fundamental limitation of the harmonic chain, hard to avoid unless the coupling is switched off when the angles reach their optimal values.

We have checked this validity condition up to a chain of eight oscillators for different combinations $\{r_l\}$, see for example figure \ref{optang}.
 Indeed, only when we approach $\frac{c}{4\omega_0^2}\sum_le^{2r_l}\gg \sum_le^{-2r_l}$ they cease to be valid.

\subsection{Optimal angles at high squeezings and strong coupling}
We have checked numerically that by increasing the quantity $\frac{c}{4\omega_0^2}\sum_le^{2r_l}$ the optimal angular relations get restricted to a smaller measure.
 This happens for the opposite regime, i.e. $\frac{c}{4\omega_0^2}\sum_le^{2r_l}\gg \sum_le^{-2r_l}$.
 The angular relations \eqref{oldAngles1},\eqref{oldAngles2} still hold, but they are restricted to areas near the values $\theta_n=n\pi/2$,
 with $n_{odd}$ even and $n_{even}$ odd. The evolution from one regime to the other can be seen in a naive representation
 in figure \ref{optang} (naive in the sense that we are equating all angles with the same parity and all squeezings).

In the case of small coupling $c/\omega_0^2\ll1$, but high squeezing, we can obtain the maximum attainable logarithmic negativity by assuming the new optimal angles.
We find the same maximum entanglement $E_N^{max}$ as in Eq.\eqref{eq8}. This means that this expression is {\it exact for any amount of squeezing}.
 This extends the validity of eq. \eqref{eq8} to any degree of squeezing as long as the coupling coefficient $c$ is small.

Though, when the coupling is higher and $c/\omega_0^2$ cannot be neglected, we can still obtain a perturbative expression:
\begin{equation}
\label{eq:entmax}
E_N^{max}\simeq\!-\frac{1}{2}\text{log}_2\!\left[\alpha\!\left(\frac{2}{N}\!\!\sum_{l,odd}\frac{e^{-2r_l}}{\omega_l}\right)\!\!\left(\frac{2}{N}\!\!\sum_{m,even}\omega_me^{-2r_m}\right)\!\right]\!.\!
\end{equation}
with
\begin{equation}
\label{correction}
\alpha=1-\gamma+\gamma^2+...
\end{equation}
and
\begin{equation}
\gamma=\frac{\sum_{lO,mE}\frac{\omega_m}{\omega_l}e^{-2r_l}e^{-2r_m} } {\sum_{lE,mO}\frac{\omega_m}{\omega_l}e^{2r_l}e^{2r_m}}
\end{equation}
where $lO,mE$ means a sum running over $l$ odd and $m$ even and viceversa.
This equation has been obtained from the expression of the local invariants in terms of $A_{l,m}$ when expanded in orders of $\chi=\sum_l e^{-2r_l}/N$ as a small quantity.
The correction factor $\alpha$ is smaller than 1 so it diminishes the argument in the logarithm, hence increasing the entanglement.
Thus, eq. (\ref{eq:entmax}) with $\alpha=1$ is a lower bound of the maximum entanglement which can be achieved.
The higher order corrections increase slightly the predicted maximum achievable entanglement. Also notice that approximately we have $\gamma=O(\chi^4)$ (it can be seen by taking similar squeezings
$r_l\simeq R$, so that $\chi\simeq e^{-2R}$) already, a small correction indeed.
That this expression for the maximum entanglement is so similar to the former one shouldn't be a surprise since it comprises the best possible combination of exponents.
Any other combination of angles would yield positive exponents together with the negative ones.

From the known expression for the irreversible work dissipated into each eigenmode \cite{gal2009b}, $W_{\text{diss.},l}=\omega_l \sinh^2 r_l$,
 we can conclude \cite{gal2009} that obtaining a high level of nonseparability requires an exponential investment in energetic resources.

\subsection{Transition between regimes}
In fig.\ref{optang} we show the transition between regimes. We see that in the upper left figure,
 which has $\frac{c}{4\omega_0^2}\sum_le^{2r_l}=0.09\ll\sum_le^{-2r_l}=0.82$, the optimal angular relations of the first regime hold.
 Note however that as we increase the squeezing, the entanglement begins to be slightly higher in regions near the new optimal angles,
 and finally gets highest in regions {\it quite} close to the optimal values. The second regime of optimal angles doesn't come until
 $\frac{c}{4\omega_0^2}\sum_le^{2r_l}$ is higher than $\sum_le^{-2r_l}$ (lower-right figure).
\begin{figure}
\label{optang}
\includegraphics[width=9cm]{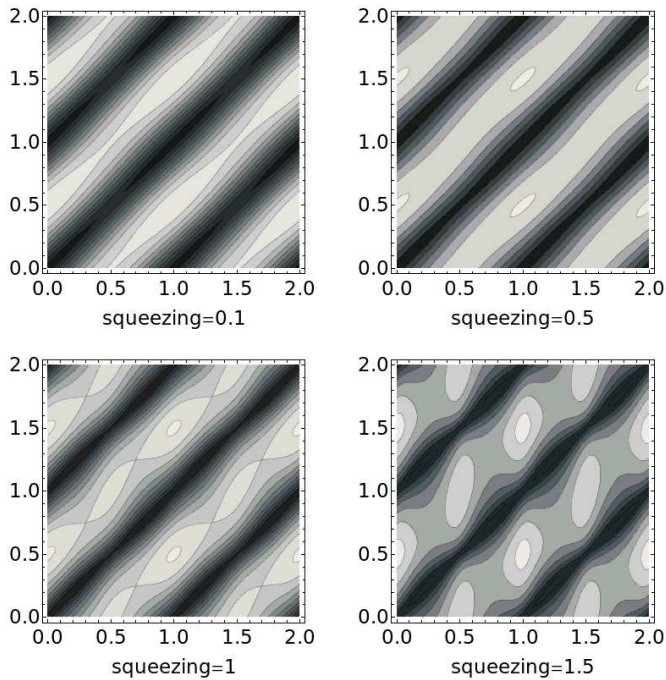}
\caption{Logarithmic negativity versus $\theta_{odd}/\pi$ (abscissa) and $\theta_{even}/\pi$ (ordinate) for equal squeezing on all eigenmodes, at $c=0.3\omega_0^2$ for eight oscillators.
The white peaks mean highest entanglement. We observe the transition between the two different regimes of optimal angles as squeezing is increased,
 with the quantities $\frac{c/\omega_0^2}{4N}\sum_le^{2r_l}=(0.09,0.2,0.55,1.5)$ and $\frac{1}{N}\sum_le^{-2r_l}=(0.82,0.37,0.14,0.05)$.}\label{optang}
\end{figure}
\begin{figure*}[t]
\includegraphics[width=15cm]{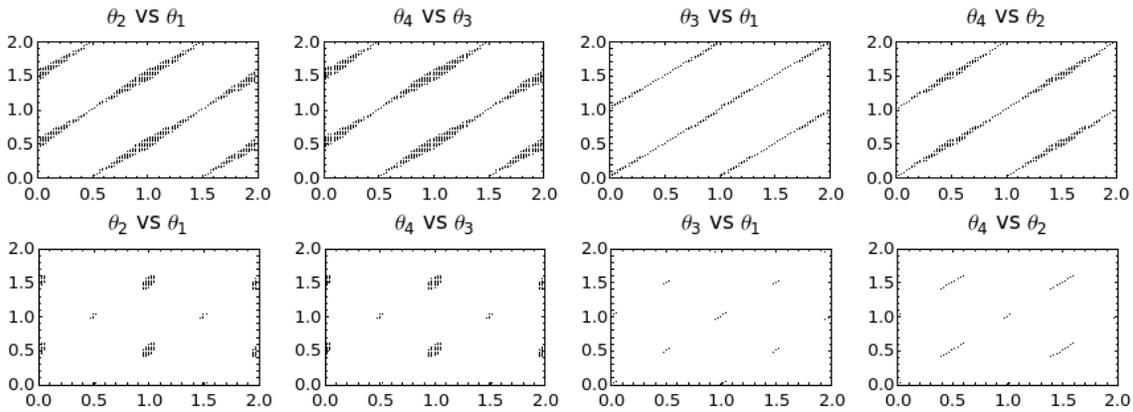}
\caption{Set of optimal angular relationships (in units of $\pi$) for $c=0.05\omega_0^2$ and $N=8$ oscillators. We have run over all angles for a set of arbitrary, but similar,
squeezings and drawn the angles with maximum entanglement (up to a deviation of $2\%$). In the upper row the set of squeezings is $(r_1=r_7,r_2=r_6,r_3=r_5,r_4)=(0.98,1.07,0.89,0.72)$
 and in the lower row $(1.92,2.26,1.90,2.37)$. The same behavior as in in figure \ref{optang} can be seen, tough in this figure we can see explicitly the optimal relations between
 angles of different {\it and} equal parity. The lower row shows that for high squeezing the region for optimal angles shrinks around the optimal values.
 The self-consistency condition yields $\frac{c}{4\omega_0^2}\sum_le^{2r_l}=0.6(6)$ and $\sum_le^{-2r_l}=2.1(1.1)$ for upper(lower) rows, showing that the condition correctly predicts the change between regimes.}
\label{optangBis}
\end{figure*}
A more thorough analysis for different combinations of squeezings $\{r_l\}$, and independent values for the angles confirms the new restricted angular relations
 which are centered around the optimal values. This can be seen in the example of figure \ref{optangBis}. There we have run over the whole space of angles for an arbitrary set of squeezings (fixed around a typical value of a given size) and only those points with highest entanglement have been drawn (up to a deviation of $2\%$). In the first row we see the regime of small squeezing, and in the second raw we show the opposite regime. It is clearly seen that the new angular relations are the same as before, but restricted to a smaller area, as explained before. The same occurs if we increase $c/\omega_0^2$ instead of/and the squeezings.

{\it Notice} that since every eigenmode's angle evolves with its own eigenfrequency, the evolution of entanglement will run over the angular space represented in figure \ref{optang}. When the area of highest entanglement gets smaller, the time spent in this region will be shorter, thus producing a more oscillatory character in the time evolution of entanglement. Concluding, entanglement can be made to increase, but it will necessarily oscillate more strongly.

\subsection{Sudden switch}
The sudden switch is a great example in order to fully analyze the attainable entanglement, since we know exactly the amount of squeezing it produces.
 An instantaneous change of the coupling from 0 to $c$, changes the eigenfrequencies from $\omega_0$ to $\omega_l=\sqrt{\omega_0^2+4c\sin^2(\pi l/N)}$.
 Such jump produces a known amount of squeezing, which according to our definitions of the variances is
\begin{equation}
\label{squeezing}
r_l=\frac{1}{2}\ln\frac{\omega_l}{\omega_0}.
\end{equation}
If we use this equality to express the logarithmic negativity between opposite oscillators, we realize that we have reduced the set of parameter to $\{\theta_l\}$ and $c$.
 Eq. \eqref{eq:entmax} yields the maximum logarithmic negativity which can be achieved for opposite oscillators:
\begin{equation}
E_N^{max.}=-\frac{1}{2}\log_2 \frac{2\alpha\omega_0^2}{N}\sum_{lO}\frac{1}{\omega_l^2}
\end{equation}
that for e.g. eight oscillators yields
\begin{equation}
E_N^{max.}=-\frac{1}{2}\log_2\frac{1+2c/\omega_0^2}{1+2c/\omega_0^2(2+c/\omega_0^2)}.
\end{equation}

This expression yields $E_N^{max.}\simeq\frac{c/\omega_0^2}{\ln2}$ for small $c/\omega_0^2$ and $E_N^{max.}\simeq\frac{1}{2}\log_2c/\omega_0^2$ for big $c/\omega_0^2$.
Clearly, by increasing the value of the coupling constant, there is no limit to the entanglement we can produce between opposite oscillators, though it must be noted that the energy cost grows linear with $c/\omega_0^2$.
By using an optimized ramp (see e.g. \cite{gal2009}) we can though produce very much entanglement without the need to reach the strong coupling regime.

\section{Entanglement propagation}
In Ref. \cite{gal2009} the optimal coupling modulation was obtained for a chain of 8 oscillators, showing maximal entanglement production between all pairs of opposite oscillators.
 In this section we would like to remark the connection between this optimization and the propagation properties of the harmonic chain.
 In particular we will see that the optimal modulation produces a propagation of entanglement along the chain without entangling oscillators,
 except for opposite ones. This is in stark contrast with the case of a sudden switch of the coupling \cite{eis04}, where the entanglement 'wave' sequentially entangles all pairs during its propagation.

\subsection{Non modulated chain}
We consider here a similar argument as given by Cubitt and Cirac \cite{cirac2008},
 which states that entanglement can be seen as correlations traveling along the chain, the properties of which can be controlled through its dispersion relation. Such dispersion relation is given by the spectrum of the chain, as we show next. 

Consider the chain in its uncoupled ground state, and a {\it sudden switch} of the coupling at $t=0$. Taking the thermodynamic limit $N\to\infty$ and $2\pi k/N\to\phi$, we have the continuous spectrum of eigenfrequencies $\omega(\phi)=[\omega_0^2+2c(1-\cos\phi)]^{1/2}$. If we take for example the correlation $\langle q_nq_m\rangle$ as representative, we obtain from eqs. \eqref{eq4},\eqref{eq5} and \eqref{squeezing}:
\begin{equation}
\langle q_nq_m\rangle=\frac{1}{8\pi\omega_0}\sum_{s=\pm 1}\int_0^{2\pi}d\phi A(\phi)\cos[\phi x+2s\omega(\phi) t]
\end{equation}
with $x=n-m\neq0$, $A(\phi)=(\omega_0^2-\omega^2(\phi))/\omega^2(\phi)$ and we have removed a constant term. This is clearly the sum of two counterpropagating wave packets, with dispersion relation given by the spectrum $\omega(\phi)$. Thus, the excitations created by the sudden switch will travel along the chain following the dispersion characteristics of the medium, given by $\omega(\phi)$. From this relation, the group velocity yields information on the dispersion characteristics of the medium. The group velocity:
\begin{equation}
\label{groupvelocity}
\frac{d\omega(\phi)}{d\phi}=\frac{c\sin\phi}{\omega(\phi)}
\end{equation}
is only rather flat when $c/\omega_0^2\to 0$. Thus, each wave with different momentum will travel with a different velocity, causing dispersion of the wave packet. Though this picture might seem too naive, it is rather powerful. It can be shown that in a more general situation where the coupling has been strongly modulated during $t<0$, but is kept constant afterwards ($t\geq 0$), the propagation of wave packets is also ruled by the dispersion relation $\omega(\phi)$.

\begin{figure}
\includegraphics[width=9cm]{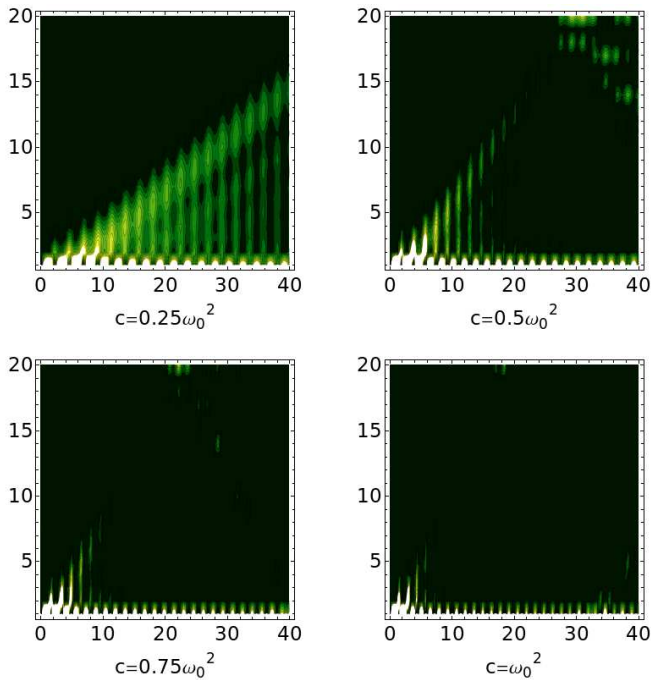}
\caption{Time (abscissa) evolution of entanglement between oscillators separated by a distance $n-m$ (ordinate) in a chain of $N=40$ oscillators after a sudden switch of the coupling. Due to translation invariance, entanglement is only a function of the distance between oscillators. The contour plot ranges from $E_N=0$ (dark) and $E_N=0.1$ (light) . We can see an increase of propagation velocities with higher $c$, but also an increase in the signal's loss. }
\label{fig3}
\end{figure}

In Ref. \cite{cirac2008} they were able to engineer the parameters so as to have nondispersive wave packets, even in a medium with high loss/gain (nonnegligible anisotropy $\gamma$). It was so because they had an extra degree of freedom, they could manipulate the coupling strength {\it and} anisotropy. In our case, the interaction Hamiltonian written in terms of (bosonic) excitations is
\begin{equation}
x_ix_{i+1}\propto a_ia_{i+1}^\dagger+a_ia_{i+1}+h.c.
\end{equation}
whereas in their case, the equivalent interaction Hamiltonian is
\begin{equation}
\label{espines}
(1+\gamma)\sigma^x_i\sigma^x_{i+1}+(1-\gamma)\sigma^y_i\sigma^y_{i+1}\propto a_ia_{i+1}^\dagger+\gamma a_ia_{i+1}+h.c.
\end{equation}
This extra parameter $\gamma$ allows for the dispersion relation to be rather flat in some cases, leading to nondispersive propagation. In our case though, 
nondispersion occurs only when $c/\omega_0^2\to 0$, which would mean no propagation at all. For finite but small coupling, fast rotating terms in interaction picture can be neglected:
\begin{equation}
a_i^{(I)}(t)a_{i+1}^{(I)}(t)+h.c.=e^{-2i\omega_0t}a_ia_{i+1}+h.c.
\end{equation}
so there is no loss/gain. However propagation will be dispersive unless an interaction of the form $(1+\gamma)x_ix_{i+1}+(1-\gamma)p_ip_{i+1}$ is used. In figure \ref{fig3} we see the propagation of entanglement after a sudden switch of the coupling. For small coupling the packet has no loss, but is highly dispersive, while higher couplings increase loss during propagation. 

\subsection{Modulated chain}
\begin{figure}[t]
\includegraphics[width=8.4cm]{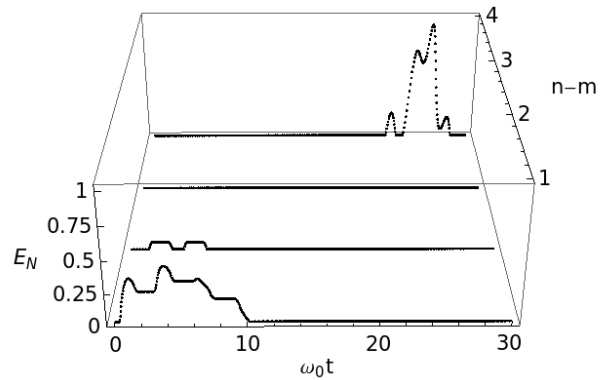}
\caption{Propagation of entanglement for an optimal modulation of the coupling (see Appendix \ref{App:OCT}) in a chain of $N=8$ oscillators as a function of time. The coupling has been switched from $c=0$ to $c=0.5\omega_0^2$. The quantity $n-m$ is the distance between oscillators. Only opposite oscillators (and slightly also the immediate neighbors) get entangled, i.e. the wave seems to get dispersed but manages to arrive to the end in its full form.}
\label{propOCT}
\end{figure}

The situation changes drastically when propagation occurs during modulation of the coupling.
We would like to emphasize that the optimal modulation in \cite{gal2009} (see fig. \ref{propOCT}) is very similar to the last figure in Fig. \ref{fig3} (highest coupling), in the sense that the entanglement wave seems to be completely destroyed along propagation. However, the wave revives completely when it reaches the end of the chain, so only entanglement between opposite oscillators is achieved.
The optimization procedure surely has to do with improving the time evolution (in interaction picture) of the loss/gain terms, which are also responsible for the creation of squeezing, such that the entanglement is created and delivered only to oscillators sitting in opposite positions. This exclusivity comes from the fact that we have maximized the functional in eq. (\ref{eq8}), designed to be the maximum achievable entanglement {\it for opposite oscillators}. Had we wanted to obtain maximal entanglement between any other pair, we would have gotten a different expression, with a different combination of exponents. Maximizing that other expression we would have come to a propagation in which entanglement is only delivered to the wanted pairs, if such an optimal modulation exists at all (remember that the case of opposite oscillators is highly symmetric, and it might be the case that we cannot deliver entanglement to nonopposite oscillators without entangling some other pairs).
Basically, changing the functional to optimize equates to changing the weight given to the different $r_l$, thus favoring delivery (once modulation has stopped and the coupling is kept constant) of the 'sum of squeezings' to pairs of oscillators with a given selected distance $n-m$.

The role of the loss/gain terms, which destroy/create excitations, in the production of distant entanglement is clearly highlighted e.g. for a driven anisotropic XY spin chain \cite{galve2009c}, where a resonant modulation of the coupling is equivalent to the limit $\gamma\to\infty$ in \eqref{espines}. That is, it maximizes the presence of creation terms in interaction picture.

\subsection{Transmission speed}
\begin{figure}
\includegraphics[width=7.5cm]{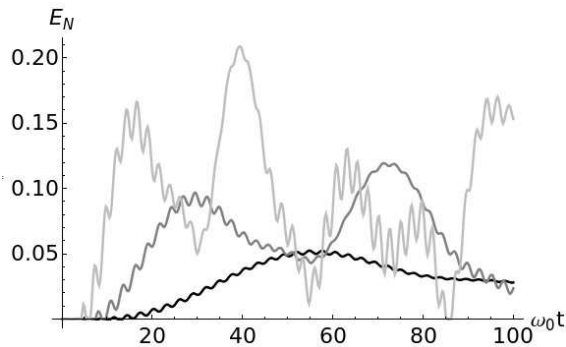}
\caption{Entanglement between opposite oscillators when the coupling is suddenly switched from 0 to $c=0.05\omega_0^2$(black), $c=0.1\omega_0^2$(gray) and $c=0.2\omega_0^2$(light gray). A speed up of the synchronization process by a factor $\sim3.5$ is observed.}\label{fig4}
\end{figure}

Finally let us comment on the validity of the group velocity eq.\eqref{groupvelocity}. If we consider the simple picture in which the highest group velocity is the one responsible for distant oscillators to start becoming entangled, we can restrict the analysis to 
\begin{equation}
v_{\rm max}=v(\phi=\pi/2)=\frac{c}{\sqrt{\omega_0^2+2c}}.
\end{equation}
So propagation can be sped up just by increasing the coupling.
In figure \ref{fig4} we show the case of a sudden switch in the coupling, where $c/\omega_0^2$ has been increased from 0.05 to 0.2;
the arrival time of the first peak of entanglement is a factor $\sim 3.5$ smaller for $c=0.2\omega_0^2$ than for $c=0.05\omega_0^2$.
Using our expression for the group velocity we obtain a factor $3.54$ . On the other hand, according to the detailed study \cite{njp} the time at which two
 oscillators at a distance $n$ are not anymore separable is $\tau\sim n\sqrt{\omega_0^2+2c}/2c $, where n is the distance between oscillators. Their expression coincides with our very simple expression coming from the group velocity $\tau=n/v_{\rm max}$.

Hence we see that reasoning in terms of a dispersive medium for excitations, correlations and entanglement yield quite similar insights. Furthermore, it allows to make quite accurate predictions from very simple arguments.

\section{Conclusion}
We have considered the limitations on the production of entanglement in a driven harmonic chain for different regimes, deriving expressions for the maximum attainable entanglement between opposite oscillators given a set of produced squeezings in the eigenmodes. The validity of those expressions  have been investigated for strong/weak coupling and moderate/high squeezing. 

We have also shown that the optimal phase relations for the eigenmodes, which provide maximum entanglement for a given set of squeezings, shrink when the squeezings increase, leading to a highly oscillatory behavior of entanglement in time. This imposes a practical limit on how much entanglement can be used in a harmonic chain, unless we are able to freeze its evolution (switching off the coupling) at its maximum value. The validity of the phase relations for weak coupling and moderate squeezing has been estimated (see Appendix \ref{App:Validity}). We have exemplified the transition to the smaller set of optimal phase relations in figs. \ref{optang} and \ref{optangBis}.

Finally, we have investigated the role of loss/gain terms in the interaction Hamiltonian, relating their effect to the validity of the RWA and the strength of the coupling. These terms are negligible for weak coupling, leading to lossless (though dispersive) propagation. It is precisely these terms, which also create excitations in the chain, that are favored by an optimization of a time dependent coupling. This creates squeezing and restricts delivery of entanglement to the selected pair of oscillators. The concept of a chain as a dispersive medium for correlations has proven fruitful for understanding the propagation characteristics, which we have used to derive in a simple fashion the time it takes for distant oscillators to become entangled. 

This work was partially supported by  the Emmy Noether Program of the DFG (Contract LU1382/1-1) and the cluster of excellence Nanosystems Initiative Munich (NIM), and partially by FISICOS(FIS2007-60327), CoQuSys(200450E566) and DiSQuC(AAEE0113/09) projects. We thankfully acknowledge R. Zambrini for manuscript revision.

\appendix
\section{Validity regime for optimal angles}
\label{App:Validity}
We will follow here a self-consistency approach in order to deduce the regime of validity of the optimal angles given in section III. We saw that for low squeezing the full relations:
\begin{eqnarray}
\theta_l-\theta_m&=&2k\pi/2 \text{\ \ \ \ \ \ \ \ \   ,\ \  $l+m$ even}\\
\theta_l-\theta_m&=&(2k+1)\pi/2 \text{\ \ ,\ \  $l+m$ odd}
\end{eqnarray}
with $k\in\mathbb{Z}$, are optimal, while for high squeezing the range of optimal angles is reduced to the smaller set:
\begin{equation}
\label{appeq2}
\theta_n=n\pi/2\text{ with $n_{odd}$ even and $n_{even}$ odd}
\end{equation}
The first relations will not hold if the squeezing and/or coupling is increased. The angular relations which do not overlap with regions near the angles in eq. \eqref{appeq2}, for example the angles $\theta_s=\pi/4$(mod $\pi$) ($s$ even) and $\theta_s=3\pi/4$(mod $\pi$) ($s$ odd), will be the first to become nonoptimal (see fig. \ref{optang}). Our strategy will be to compare the symplectic eigenvalue in \eqref{eq_eigen} for the latter angles and see when they give a higher eigenvalue than the one given by the actually optimal angular values in \eqref{appeq2} (i.e. a lower attainable entanglement). For the first regime we choose $\theta_{even}=\pi/4$ and $\theta_{odd}=3\pi/4$, while for the other regime we choose $\theta_{even}=\pi/2$ and $\theta_{odd}=0$ (note that this retains the generality of the argument). Following the notation of Eq. \eqref{eq_eigen} we can write:
\begin{eqnarray}
x^{(\frac{\pi}{4},\frac{3\pi}{4})}&=&\sum_{lO,mE}(1+\frac{c}{2\omega_0^2})e^{2(r_l+r_m)}\\
y^{(\frac{\pi}{4},\frac{3\pi}{4})}&=&\frac{1}{4}\left(\sum_{lO}(1+\frac{c}{2\omega_0^2})e^{2r_l}\right)^2\times\nonumber\\
&\ &\left(\sum_{mO}e^{-2r_m}+\frac{c}{2\omega_0^2}e^{2r_m}  \right)^2
\end{eqnarray}
and
\begin{eqnarray}
x^{(\frac{\pi}{2},0)}=&(1+\frac{c}{\omega_0^2})\sum_{lE,mO}e^{2(r_l+r_m)}&\\
y^{(\frac{\pi}{2},0)}=&(1+\frac{2c}{\omega_0^2})\left(\sum_{lO}e^{-2r_l}\right)\left(\sum_{mO}e^{2r_m}\right)\times&\\
&\left(\sum_{lE}e^{2r_l}\right)\left(\sum_{mE}e^{-2r_m}\right),&
\end{eqnarray}
where we have approximated $\frac{\omega_m}{\omega_l}\simeq(1+\alpha_{l,m}\frac{c}{\omega_0^2})$ (that is, $c/\omega_0^2\ll1$) and taken $\alpha_{l,m}$ of order 1. We use also the previous notation where $lO,mE$ mean $l$ over odd values and $m$ running over even values.

The mixture of sums for odd, even values of these terms makes it difficult to obtain a clear picture. Therefore we employ a further assumption, that the squeezings are similar for even-odd eigenmodes in the sense that if we define 
\begin{equation}
B_{O,(E)}^\pm:=\sum_{lO(E)}e^{\pm 2r_l}
\end{equation}
we approximate
\begin{equation}
B_O^\pm\sim B_E^\pm\equiv B^\pm\ .
\end{equation}
This is reasonable in the sense that $c(t)$ is the only controllable parameter in the system, which squeezes all eigenmodes even if it is tailored to only squeeze a given parity. We further notice that $B^-/B^+\ll1$ whenever the squeezings are relative strong (e.g. $r\gtrsim 1/2$).

With these assumptions we can now compare the symplectic eigenvalues of each regime to conclude that
\begin{equation}
|\nu_-^{(\frac{\pi}{4},\frac{3\pi}{4})}|>|\nu_-^{(\frac{\pi}{2},0)}|\iff \frac{c}{4\omega_0^2}B^+>B^-
\end{equation}
, i.e. when
\begin{equation}
\frac{c}{4\omega_0^2}\sum_le^{2r_l}>\sum_le^{-2r_l}\ .
\end{equation}

We have checked, up to a chain of eight oscillators, that the optimal angles are valid, according to the above relation, for different combinations $\{r_s\}$.
 They cease to be valid only when we approach $\frac{c}{4\omega_0^2}\sum_le^{2r_l}\gg \sum_le^{-2r_l}$.

\section{Optimal control algorithm}
\label{App:OCT}
Optimal control theory (OCT) is a branch of applied mathematics whose objective is the determination of the function which optimizes a given cost functional \cite{kir70}. A very famous particular case leads to the Euler-Lagrange equations of motion in classical mechanics, when the action functional is extremized with respect to trajectories. We derive the OCT algorithm which has been used here and in \cite{gal2009,gal2009b}, for the sake of reproducibility.

In our particular case, we want to find the coupling modulation $c(t)$ which maximizes equation \eqref{eq:entmax}. In this problem, this can be achieved with purely classical equations. The solution to the time evolution of the chain is fully determined by the equations of motion of the normal modes, which are quantum harmonic oscillators with frequency modulation. Examining eqs. \eqref{eq5} we observe that the average energy of a given normal mode is 
\begin{equation}
\langle E_s\rangle=\frac{\omega_s}{2}\cosh2r_s
\end{equation}
therefore the squeezing of a given mode can be expressed purely by means of the quantity $2\langle E_s\rangle/\omega_s$, or in terms of $\langle p^2\rangle$ and $\langle x^2\rangle$. Thus, for a given modulation of the coupling coefficient $c(t)$ the squeezing of eigenmode $s$ can be obtained from the {\it classical} equation of motion of a frequency modulated oscillator with frequency $\sqrt{\omega_0^2+4c(t)\sin^2(\pi s/N)}$ and arbitrary initial conditions, provided that its initial energy is equal to the ground state energy of the quantum oscillator.

The aim of the optimal control algorithm is to minimize the cost functional $J\left(c(t)\right)$ of the coupling coefficient in equation \eqref{eq8}. In our case this functional depends only on the coordinates of the modulated classical oscillator at the total modulation time $\tau$, hence $J\left(c(t)\right)=h\left(\vec{x}(\tau)\right)$. The vector $\vec{x}=(\vec{x}_1,...,\vec{x}_N)$ contains the positions and momenta of the eigenmodes such that $\vec{x}_i=(x_i,p_i)$. A simple way to impose the equations of motion of the eigenmodes is to include Lagrange multipliers in the cost functional, so at the end we have
\begin{equation}
h\left(\vec{x}(\tau)\right)=\int_0^\tau dt\left\{\frac{\partial h}{\partial\vec{x}}\dot{\vec{x}}+\vec{\xi}\left[\vec{a}(\vec{x},c)-\dot{\vec{x}}\right]\right\}
\end{equation}
where $\vec{\xi}=(\vec{\xi}_1,...,\vec{\xi}_N)$ are the multipliers (with $\vec{\xi}_i=(x_i^\xi,p_i^\xi)$ ), or costates in optimal control jargon, and the equations of motion of the eigenmodes are simply $\dot{\vec{x}}=\vec{a}(\vec{x},c)$. In order to find the equations governing the optimal control algorithm, we need to follow an approach very similar to the deduction of the Euler-Lagrange equations, which can be found in\cite{kir70}. The resulting equations are:
\begin{eqnarray}
1)\ & \dot{\vec{x}}_i=&\vec{a}_i(\vec{x},c)\nonumber\\
2)\ & \dot{\vec{\xi}}_i=&-\frac{\partial(\vec{\xi}\cdot\vec{a})}{\partial\vec{x}_i}\nonumber\\
3)\ & \vec{\xi}(\tau)=& \left. \frac{\partial h}{\partial\vec{x}} \right| _\tau\nonumber\\
4)\ & \vec{\xi}^*\cdot\vec{a}(\vec{x}^*,c^*)\leq&\vec{\xi}^*\cdot\vec{a}(\vec{x}^*,c)\ ,\ \forall\ c\nonumber
\end{eqnarray}
where 1) is the evolution equation of the eigenmodes, 2) is the evolution equation of the costates, 3) are the initial conditions of the costates (note that they are stated at the final time) and 4) is Pontryagin's minimum principle which states that {\it the optimal trajectories are those which minimize $\vec{\xi}\cdot\vec{a}$ all along the trajectory} and the stars mean "optimized" (it should be noted here that Pontryagin's minimum principle is an extension of optimal control theory, typically used when the control parameter $c(t)$ is bound in a given range). The fact that the costates have "initial conditions" at the final time prevents us from obtaining the optimal $c(t)$ in one go. There are several alternatives for solving the problem, but we have chosen to use an iterative scheme with steepest descent as follows:\\
a)Choose a trial function $c(t)$.\\
b)Evolve the eigenmodes until $\tau$ (and record their trajectory).\\
c)Obtain, through equations 3), the initial conditions for the costates.\\
d)Evolve backwards in time the costate through their equations 2) (and record their trajectory).\\
d)Change the old $c(t)$ by the amount $\alpha\partial(\vec{\xi}\cdot\vec{a})/\partial c$.\\
e)Repeat the process starting with the improved $c(t)$.\\
The factor $\alpha$ has to be changed according to the performance of the algorithm. Too big an $\alpha$ will tend to find too fast the solution, which we will recognize as an oscillating solution. If $\alpha$ is too small, the convergence will be slow.

In our system the equations of motion for the eigenmodes are $\dot{x}_i=p_i$, $\dot{p}_i=-\omega_i^2x_i$, while the equations for the costates are $\dot{x}_i^\xi=\omega_i^2p_i^\xi$ and $\dot{p}_i^\xi=-x_i^\xi$. We also have $\vec{a}=(p,-\omega^2x)$ and thus the gradient function $\partial(\vec{\xi}\cdot\vec{a})/\partial c=-4\sum_k\sin^2\left(\frac{\pi k}{N}\right)x_ip_i^\xi$. The cost functional can be chosen to be the argument in the logarithm in eq. \eqref{eq:entmax}, the inverse of the sum of energies, etc. A bit of trial and error is needed until a suitable cost functional is found.

A special comment should be made here. It must be noticed that the role of the cost functional $h$ in the improvement in $c(t)$ is quite hidden in the algorithm. Its effect is felt only through the initial values for the evolution of the costates, which at the end affect the gradient function $\partial(\vec{\xi}\cdot\vec{a})/\partial c$.

\end{document}